\documentclass[twocolumn,eqsecnum,aps]{revtex4} 
\def\be{\begin{equation}} 
\def\ee{\end{equation}} 
\def\bea{\begin{eqnarray}} 
\def\eea{\end{eqnarray}} 
\def\eps{\varepsilon}

\begin{document} 
\title{Chiral Condensate and Short-Time Evolution  of  QCD$_{1+1}$ on the Light-Cone}  
\author{Matthias Burkardt} 
\address{Department of Physics\\ 
New Mexico State University\\ 
Las Cruces, NM 88003-0001\\U.S.A.} 
\author{Frieder Lenz and Michael Thies} 
\address{Institut f\"ur Theoretische Physik III\\ 
Universit\"at Erlangen-N\"urnberg\\ 
Staudtstra\ss e 7\\ 
D-91058 Erlangen\\ 
Germany} 
\date{\today} 
\begin{abstract}
Chiral condensates  in the trivial light-cone  
vacuum emerge if defined as short-time limits of fermion propagators.
In gauge theories, the necessary inclusion of a gauge string in combination with the characteristic 
light-cone infrared singularities contain the relevant non-perturbative ingredients responsible for formation of
the condensate,
as demonstrated for the 't Hooft model.
\end{abstract} 
\maketitle 

\section{Introduction} 
The triviality of the light-cone vacuum is the origin of most of the 
simple properties of field theories if quantized on the light-cone (for a recent review, see  
Ref. \cite{BPP98}). For purely kinematical reasons 
there is no distinction between the ground state of free and 
interacting theories. In view of the many physical consequences 
usually attributed to non-trivial vacua, this has raised considerable 
concern about the equivalence of light-cone and equal-time quantization 
schemes. While for bosonic theories the common belief is that the 
key lies in the constrained zero-mode dynamics \cite{bosons}, 
in fermionic theories a way out of this apparent contradiction has been  
sought in prescriptions for  regularization of the divergent condensates such as the ``parity invariant" regularization relating
 IR and UV cutoffs \cite{parity}. It is doubtful that universal, kinematical prescriptions exist which describe the formation of
 condensates in theories such as gauge theories where chiral symmetry breaking is not tantamount  
to fermion mass creation.
In the present work, the proposal \cite{LTY00} to define order parameters as vacuum expectation 
values of  products of Heisenberg operators, infinitesimally split in light-cone 
time direction (in addition to a space direction) will be shown to yield the correct condensate in QCD$_{1+1}$ --- the 't Hooft
 model \cite{thooft}. Unlike in standard quantization, the short-time limit of Heisenberg operators differs non-perturbatively 
from the corresponding Schr\"odinger operators. This connection between condensates and the non-triviality of the singular
 behavior of correlation functions at short light-cone times will be the focus of this work. 

The condensate of the 't Hooft model
 has been evaluated in \cite{zhit,burk96}. These studies make use of general relations between properties
 of the excited states
and the condensate, derived in standard quantization
 (Oakes-Renner relation and sum rules). Here we will present
 a direct calculation of the condensate which makes use explicitly of the triviality of the light-cone
 vacuum on the one hand and
 the non-perturbative nature of the short-time limit of light-cone correlation functions on the other.       

\section{Condensate and Short-Time Evolution} 
We define the condensate with respect to that of the non-interacting theory,
\begin{equation}
  \label{sub}
\langle \bar{\psi}\psi \rangle=\lim_{\eps \to 0}  \left[\langle \bar{\psi}\psi \rangle_{\eps}-\langle \bar{\psi}\psi 
\rangle_{\eps}^{0}\right]  \ ,
\end{equation}
with
\be 
 \langle \bar{\psi}\psi \rangle_{\eps}
=\langle 0 | \bar{\psi}(\eps) P{\rm e}^{{\rm i}g \int_0^\eps {\rm d}x^\mu A_\mu} 
\psi(0)|0\rangle 
\label{eq:cond} 
\ee
regularized in a gauge invariant way. 
For evaluating the point-split condensate (\ref{eq:cond}), 
we first consider a generic matrix element of the type 
\be 
M(\eps)=\langle 0 |A(\eps) P {\rm e}^{{\rm i}g \int_0^s {\rm d}s' 
\frac{{\rm d}x^{\mu}}{{\rm d}s'}A_{\mu}(x(s'))}B(0)|0\rangle 
\label{n1} 
\ee 
where $x(s)$ is a straight path with $x(0)=0$ and 
$x(s)=\eps$. By shifting  the argument 
of the gauge field $A_{\mu}$ and using translational 
invariance of the vacuum, we can represent $M(\eps)$ 
as 
\be 
M(\eps)=\langle 0 | A(0) W(s) B(0)| 0 \rangle 
\label{n3} 
\ee 
with the point splitting 
now specified by the operator $W(s)$  (momentum operator $P_{\mu}$), 
\be 
W(s)={\rm e}^{-{\rm i} \eps^\mu  P_\mu} P  {\rm e}^{{\rm i}g \int_0^s {\rm d}s' 
\frac{{\rm d}x^\mu}{{\rm d}s'}A_{\mu}(x(s'))}\ . 
\label{n4} 
\ee 
Differentiation with respect to $s$,
\bea 
\frac{{\rm d}W(s)}{{\rm d}s} &=&  -{\rm i}\frac{{\rm d}x^\mu}{{\rm d}s}\left(P_\mu -gA_\mu (0)\right)W(s) \ ,
\label{eq:diff} 
\eea 
and integration of this  differential equation with 
the initial condition $W(0)=1$  
allows one to simplify this expression, yielding
\be 
W(s)={\rm e}^{-{\rm i}\eps^\mu \left(P_\mu-gA_\mu(0)\right)}. 
\label{ws}
\ee 
With this form of the operator $W$ the chiral 
condensate of Eq.
(\ref{eq:cond}) is seen to be given by the space-time evolution of a system of light
 quarks coupled to the current of an infinitely heavy quark.
Thus, on the light-cone with its kinematical vacuum,  the dynamics of a  heavy-light quark system 
determines the chiral condensate.  
So far, everything is rather general and applies equally well to QCD in 4 dimensions. 
 We now specialize  to the 't Hooft model  \cite{thooft} and write 
 Eq. (\ref{eq:cond}) as  \be 
 \langle \bar{\psi}\psi \rangle_{\eps}
=\langle 0 | \bar{\psi}_i(0) ( {\rm e}^{ 
-{\rm i}\eps^+ H_{\rm eff}})_{ij} \psi_j(0) |0\rangle  
\label{n10} 
\ee 
with the effective Hamiltonian in light-cone gauge  ($A_- =0$),
\be 
(H_{\rm eff})_{ij}=(P_+ +  \lambda P_-)\delta_{ij}  - g[A_+(0)]_{ij} .  
\label{n10a} 
\ee 
We have 
denoted the slope of the path, $\eps^-/\eps^+$, by $\lambda$ and 
will take $\eps^+$ and $\lambda$ as independent parameters 
from now on; due to the subtraction of the free value, the condensate  will turn out to be independent of  $\lambda$.   
$P_+$ and $P_-$ are the Hamiltonian 
and momentum operator for the 't~Hooft model, respectively. 

For evaluation of the chiral condensate, we represent the spinor $\psi$ in terms of 
the unconstrained right-handed component  $\varphi$,  
\be 
\psi(x)=\frac{1}{2^{1/4}} \left( \begin{array}{c} 1\\ \frac{m}{{\rm i}\sqrt{2}\partial_{-}}   \end{array} 
\right) \varphi  \ ,
\label{n11} 
\ee 
and find (after Fourier transforming the fermion fields) 
\be 
 \langle \bar{\psi}\psi \rangle_{\eps}
= \int \frac{{\rm d}p}{2\pi} \frac{m}{p} C(p,\eps^+) 
\label{n13} 
\ee 
with 
\be 
C(p,t) = \int \frac{{\rm d}q}{2\pi} \langle 0 | \varphi_i^{\dagger}(p) 
( {\rm e}^{-{\rm i}  H_{\rm eff}\, t})_{ij} \varphi_j(q) | 0 \rangle \ . 
\label{n14} 
\ee 
In order to compute this correlation function we derive
 its equation of motion. In    
\be 
{\rm i}\dot{C}(p,t)= 
\int \frac{{\rm d}q}{2\pi} 
\langle 0 | \varphi_i^\dagger(p)\left(H_{\rm eff} 
 {\rm e}^{-{\rm i} H_{\rm eff}\, t}\right)_{ij} \varphi_j(q) 
|0\rangle \ , 
\label{n15}  
\ee 
we treat the two terms of
$H_{\rm eff}$  (cf. Eq.  (\ref{n10a}))
separately.  In the first term, 
  the product  of the operators $(P_+ +\lambda P_-)$ and $\varphi_i^\dagger(p)$ can
be replaced by their 
commutator.  In the large $N$ limit, this 
commutator  generates a combination of the quark self-energy and momentum 
where, following 't~Hooft's original paper \cite{thooft},  
\be 
\langle 0| [\varphi_i^\dagger(p), P_+ +\lambda P_-] =  \left( \frac{m_r^2}{2p}+ 
\lambda p \right)\langle 0 | \varphi_i^\dagger (p) \ , 
\label{n16} 
\ee 
with the self-energy given in  principal value prescription by 
\be 
m_r^2\equiv m^2-\frac{Ng^2}{2\pi}\ . 
\ee 
The right hand side of (\ref{n16})  combines with the remainder 
of the matrix element in (\ref{n15}) to yield again $C(p,t)$. 
The  term proportional to $A_+(0)$ in $H_{\rm eff}$ can be expressed via 
the Poisson equation  
in terms of the fermion color charge density,
\be 
[A_+(0)]_{ij} = -\frac{g}{2} \int \!\!\!\!\!\! - \frac{{\rm d}p'}{2\pi} \frac{{\rm d}p''}{2\pi} 
\frac{ \varphi_j^\dagger(p')\varphi_i(p'')}{(p'-p'')^2}\  . 
\label{n20} 
\ee 
With this result, Eq. (\ref{n15}) can be simplified  by replacing in the large 
$N$ limit   the operator
\be 
\varphi_i^{\dagger}(p)\varphi_i(p') \to  2\pi \delta(p-p') N \theta(-p), 
\label{n21} 
\ee 
 by its expectation value. Here the triviality of the light-cone vacuum is explicitly used. Choosing units such that
\begin{equation}
\frac{Ng^2}{2\pi}=1 \ ,
\label{n22a}
\end{equation}
factoring out a step function $\theta(-p)$ from $C(p,t)$ and changing $p$ into $-p$,
the time evolution of $C$ can finally be cast into the form of a typical 
light-cone Schr\"odinger equation, 
\bea 
{\rm i}\dot{C}(p,t) &=& 
\left( \frac{m^2-1}{2p}+\lambda p \right)C(p,t)+ \frac{1}{2p}C(0,t)  \nonumber \\ 
&   
- &\frac{1}{2}\int_0^{\infty}\!\!\!\!\!\!\!\!\!\!\! - \ {\rm d}p' 
\frac{1}{p^{\prime}-p}\frac{\partial C(p^{\prime},t)}{\partial p^{\prime}}\ .   
\label{n23} 
\eea 
This evolution equation for $C(p,t)$ at short times together with the 
initial condition (cf. Eq. (\ref{n14}))
\be 
C(p,0)=N 
\label{n23a} 
\ee 
determines the condensate. We note that  
$$ {\rm i}\dot{C}(p,t=0) = N\left( \frac{m^2}{2p}+\lambda p \right) .$$
For non-interacting fermions,
\begin{equation}
  \label{Cfree}
  C_{0}(p,t) =N \exp\left\{- {\rm i}t\left(\frac{m^2}{2p}+\lambda p\right) \right\} 
\end{equation}
solves the evolution equation with the correct initial condition.
Due to the presence of singularities, 
$C(p,t)$ deviates significantly for arbitrarily small times from its 
initial value $N$. 
Characteristic for the short-time light-cone dynamics is the infrared singularity
which implies 
\be 
\lim_{t \to 0}C(0,t)=0\neq  \lim_{p\to 0}C(p ,0) \ .
\ee  
The short-time behavior for large momenta is the same in the
interacting and in the free theory. It is therefore irrelevant for the 
evaluation of the condensate and we  drop in the following 
the ultraviolet regulator,
\be 
\lambda = 0\ .
\label{la0} 
\ee
In this case  the evolution 
equation together with the initial condition implies that  $C(p,t)$ (as well as $C_{0}(p,t)$) depends only on the ratio of the 
variables $p$ and $t$,  
\be
C_{(0)}(p,-{\rm i}\tau) = N K_{(0)}(p/\tau) \ ,
\label{K0}
\ee
where we have switched to imaginary time.
With Eq. (\ref{K0}) the time evolution is converted 
into the integro-differential equation 
\be
q\frac{{\rm d}K(q)}{{\rm d}q}= 
\frac{m^2-1}{2q}K(q)-\frac{1}{2}\int_0^{\infty}\!\!\!\!\!\!\!\!\!\!\!\! - \ {\rm d}q' 
\frac{1}{q^{\prime}-q}\frac{{\rm d} K(q^{\prime})}{{\rm d}q^{\prime}}   
\label{K1} 
\ee
and the asymptotic behavior of $K(q)$ is determined by the initial condition for $C(p,t)$, 
\be
\lim_{q \to \infty}K(q)=1\ .
\label{nor}
\ee
We can also determine the infrared behavior of $K(q)$. Using  results for Hilbert transforms of 
powers (cf. \cite{erd}) it is seen that for small $q$
\be
K(q)\sim q^{\beta_{0}}
\label{sq}
\ee
where $\beta_{n}$ denote the  solutions of the 't Hooft boundary condition
\begin{equation}
\pi q \cot \pi q = 1- m^2 \ 
\label{s1}
\end{equation}
ordered according to
\begin{equation}
q=\pm \beta_n \ , \quad n=0,1,2... \quad {\rm with} \quad \beta_n \in [n,n+1]\ .
\label{s2}
\end{equation} 
Thus the (confining) interaction in the 't Hooft model changes the essential singularity of 
$K_{0}(q)$ of the non-interacting theory 
$$ K_{0}(q) = {\rm e}^{-m^2/2q}$$
to a branch point. This remarkable phenomenon is due to  the presence of the gauge string
 in  the correlation function $C(p,t)$. For comparison we remark that the same calculation in the
Gross-Neveu model, involving a similar large $N$ approximation, yields the non-interacting form
 of the correlation function with the fermion mass modified by the interactions.
\newline 
For the general case, we have not been able 
to express  the solution of Eq. (\ref{K1})  via the Mellin transform (see below) 
in simple terms. In the chiral limit ($\beta_{0}\approx \sqrt{3} m /\pi\rightarrow 0$),
$$C(p,-{\rm i}\tau)\approx N\, (p/\tau)^{\beta_{0}\theta(1-(p/\tau))} .$$ 
It can be verified that  Eq. (\ref{K1}) is satisfied up to terms of O($\beta_{0}$). This expression displays
the subtleties of the $p,t \rightarrow 0$ limit. It reproduces in the chiral limit the exact value for the condensate (see below),
i.e., the condensate is directly connected to the change in the infrared singularity of the quark propagator.
\newline   
As a consequence of the light-cone singularity in the infrared, determination of the short-time 
behavior of  
$C(p,t)$ requires  a non-perturbative calculation of the function $K(q)$.
In terms of $K_{(0)}(q)$, the condensate is written  (cf. Eqs. (\ref{sub}),(\ref{n13})) as
\begin{equation}
  \label{scc}
\langle \bar{\psi}\psi \rangle= -N \frac{m}{2\pi} \int_{0}^{\infty}\frac{{\rm d}q}{q} (K(q)-K_{0}(q))  \ .
\end{equation}
Due to the scaling property (\ref{K0}) the dependence on the regulator $\eps^{+}$ 
has disappeared entirely from the  expression of the condensate.  
  \newline
\section{Calculation of the Condensate}
We now briefly sketch the evaluation of the condensate by Mellin transformation of the  equation for  $K$.
 The techniques developed in  \cite{asympt} will be used.
We define
\be
\gamma(\kappa) = \int_{0}^{\infty} {\rm d}q\, q^{\kappa-1}K(q) \ .
\label{Mt1}
\ee
The Mellin transform is only well defined if 
\be
-\beta_{0} < \kappa <0 \ .
\label{rgi}
\ee
For non-negative $\kappa$ the integral (\ref{Mt1}) diverges at large values of $q$, while the infrared
 behavior (\ref{sq}) entails the lower limit. 
The Mellin transform converts  Eq. (\ref{K1})   into the recursion relation (cf. \cite{deb})
\be
\gamma(\kappa)= -\frac{1}{2\kappa}\left[m^2-1+\pi(\kappa-1)
\cot \pi\kappa\right]\gamma(\kappa -1) \ .
\label{rec}
\ee
As can be easily verified,
\bea
& &\gamma(\kappa) = {\cal N} \left( \frac{\pi}{2} \right)^\kappa 
\frac{\exp \left\{ - 2 \pi \int_0^\kappa {\rm d}u \left(\frac{u+ 
\frac{1}{2}\sin^2 \pi u}{\sin 2\pi u}\right) \right\} 
}
{\beta_{0}\cos \pi(\kappa+\beta_0+1/2)} \nonumber\\ 
&\cdot& \prod_{n=1}^{\infty} \left( \frac{1+ \frac{(m^2-1)\tan \pi \kappa }
{\pi \beta_{n-1}}}
{1+\frac{(m^2-1)\tan\pi \kappa}{\pi(\kappa+n)} } \right)\frac{\pi\beta_{0}-(m^2-1)\tan \pi \kappa }{\pi\kappa+
(m^2-1)\tan \pi \kappa }\nonumber\\  
\label{s4}
\eea
solves the recursion relation (\ref{rec}). Condition (\ref{nor}) determines the  
normalization ${\cal N}$. 
The solution has the expected poles at the endpoints of the regularity 
interval (\ref{rgi}) and is free of singularities   within this interval.
 
According to Eq. (\ref{scc}) the condensate is directly given by the Mellin transform 
 \begin{equation}
\langle \bar{\psi} \psi \rangle = - N \frac{m}{2\pi} \lim_{\kappa \to 0} 
\left( \gamma(\kappa)-\gamma_0(\kappa)
\right) 
\label{s8}
\end{equation}
where 
\begin{equation}
\gamma_0 (\kappa) = \left( \frac{m^2}{2} \right)^\kappa \Gamma (-\kappa)  
\label{s3}
\end{equation}
is the Mellin transform of $K_0(q)$ for non-interacting fermions.  
Expansion of the Mellin transforms  (\ref{s4}, \ref{s3}) around $\kappa=0$,
with the normalization factor ${\cal N}$ chosen such that the singular pieces ($\sim 1/\kappa$) cancel, yields
\begin{eqnarray}
\langle \bar{\psi}\psi \rangle &=& -N \frac{m}{2\pi}\left\{  1+\gamma +\ln \left(\frac{m^2}{\pi}\right)  \right.  \label{s10}  \\  
&& + \left.
(1-m^2)\left[\frac{1}{\beta_0}
+\sum_{n=1}^{\infty} \left( \frac{1}{\beta_n}-\frac{1}{n}\right)\right] \right\} .\nonumber
\end{eqnarray}
This agrees with the  result obtained in \cite{burk96}.  
In particular in the chiral limit where the $1/\beta_0$ term dominates, the result 
\begin{equation}
\langle \bar{\psi} \psi \rangle =  - \frac{N}{\sqrt{12}} \ ,
\label{s12}
\end{equation}
first derived in \cite{zhit} is reproduced.  The calculations  \cite{zhit,burk96} 
 are based on low energy theorems 
and connect the condensate with properties of the mesons of the 't Hooft model (in particular the ``Goldstone
 boson" in the chiral limit).  In our approach with the quark propagator as the 
essential ingredient, the condensate is determined by the short-time behavior of a heavy-light 
quark system.  The equivalence of these quite different approaches suggests a relation  
in light-cone quantization similar to the relation  in 
ordinary coordinates  based on chiral Ward identities  which connects  
quark propagators and  the Goldstone boson Bethe-Salpeter wave function  (cf. \cite{mira93}).    
  
\section{Conclusions} 
In quantum field theory,  condensates  are calculated in general as  expectation values of Schr\"odinger operators
 in the corresponding vacua.
 Non-vanishing order parameters reflect the non-triviality of the vacuum. In light-cone quantization, the kinematical 
 structure of the
 light-cone vacuum gives rise to trivial vacuum expectation values of Schr\"odinger operators which therefore cannot
 serve as order
 parameters. On the light-cone, condensates have to be defined as vacuum expectation values of limits of Heisenberg
 operators \cite{LTY00}. In this way, the condensate is obtained as the short light-cone time limit of an appropriate
 correlation
 function. In light-cone quantization, non-vanishing order parameters reflect the non-triviality of the short-time
 limit of the relevant
 Heisenberg operators.\newline  
We have carried out a study of the chiral condensate of two dimensional  
QCD within  light-cone quantization. A direct and explicit 
calculation of the condensate within the 't Hooft model has been presented. 
The kinematical structure of the light-cone vacuum has 
been an essential ingredient in this calculation. Our calculation 
shows that the short-time limit of the quark propagator is afflicted by non-perturbative physics. On the light-cone this
 correlation 
function is singular in the infrared; the singularity depends on the dynamics. The essential 
singularity of the non-interacting theory is converted by the 
interactions in QCD$_{1+1}$ to a branch point --- a phenomenon which defies a perturbative 
description and is responsible for generation of the condensate in the chiral limit.  The resulting fermion correlation
 function differs
 significantly at short light-cone times from the correlation function of the Gross-Neveu model. In this two dimensional
 model a chiral
 condensate emerges in the process of mass generation \cite{LTY00}. Here the essential singularity in the infrared
 persists, its
 parameters are modified by interactions. The successful description of the condensates in these dynamically very
 different models
 strongly  supports the idea of reconciling the non-trivial vacuum properties with the kinematical nature of the 
light-cone vacuum by
 the dynamical, non-perturbative  short light-cone time limit of Heisenberg operators.

\vskip 0.5cm
\noindent {\bf Acknowledgments} 
M.B. was supported by a grant from DOE (FG03-95ER40965) and through
Jefferson Lab by contract DE-AC05-84ER40150 under which the Southeastern
Universities Research Association (SURA) operates the Thomas Jefferson
National Accelerator Facility.
\vskip 0.5cm 
\

\end{document}